\documentclass[a4paper,11pt]{article}
\usepackage{jcappub} % for details on the use of the package, please
\usepackage[T1]{fontenc} % if needed
\usepackage{subfigure}
\usepackage[normalem]{ulem}

\usepackage{enumitem,amssymb,amsmath}

\def \f{\frac}

\def \o{\omega}

\def \a{\alpha}
\def \b{\beta}

\def \s{\sigma}

\def \a{\alpha}

\def \th{\theta}
\def \r{\rho}
\def \c{\chi}
\def \T{\Theta}

\def \ra{\rightarrow}

\title{\boldmath Low Mass Naked Singularities from Dark Core Collapse}

\author[a]{Chandrachur Chakraborty}
\author[b]{Sudip Bhattacharyya}
\author[c]{and Pankaj S. Joshi}

\affiliation[a]{Manipal Centre for Natural Sciences, Manipal Academy of Higher Education, \\ Alevoor Road, Manipal, 576104 India}
\affiliation[b]{Department of Astronomy and Astrophysics, Tata Institute
of Fundamental Research,\\ 1 Homi Bhabha Road, Colaba, Mumbai, 400005 India}
\affiliation[c]{International Centre for Space and Cosmology, Ahmedabad University,
\\
Navrangpura, Ahmedabad, 380009 India}

\emailAdd{chandrachur.c@manipal.edu}
\emailAdd{sudip@tifr.res.in}
\emailAdd{psjcosmos@gmail.com}

\abstract{Near-solar mass black holes (BHs) could have been involved in the two recent gravitational wave events, GW190425 and GW190814. Since such a low mass BH cannot be formed via stellar evolution, a model has been proposed based on the core collapse of a neutron star initiated by a certain number of dark matter (DM) particles. In this process, the accumulated DM particles collapse to form a tiny BH inside the neutron star, and the entire neutron star is transmuted into a BH after a certain time due to the accretion of matter by the endoparasitic BH from its host. Here, we argue that, depending on the initial conditions, 
a dark core collapse could give rise to either a BH or a naked singularity.
For example, if the accumulated cloud of DM particles in the core of a neutron star can be modeled as an anisotropic fluid and it fulfils the criterion for collapse, an endoparasitic naked singularity could form instead of an endoparasitic BH. 
Immediately after its formation, the naked singularity should begin accreting matter from the host neutron star, 
thus eventually transmuting the entire host into a near-solar mass, relatively slowly-spinning naked singularity.
We also propose a general technique to constrain the DM particle--neutron scattering cross section using the lack of pulsars near the Galactic centre and assuming that these missing pulsars have already been transmuted into BHs and/or naked singularities. 
Thus, the missing pulsars also indicate the existence of many such singularities near the Galactic center.}


\begin{document}
\maketitle
\flushbottom

\section{\label{s1}Introduction}
Recent gravitational wave events GW190425 \cite{gw1} and GW190814 \cite{gw2} have suggested the existence of near-solar-mass collapsed objects which cannot be formed via stellar evolution. These two GW events have also opened up the possibility of reliable future detections of near-solar mass collapsed objects with the advanced Laser Interferometer Gravitational-Wave Observatory (aLIGO;  \cite{aligo}), Lunar Gravitational-wave Antenna (LGWA; \cite{lgwawp}) etc.
An important question of physics and astronomy is  
how such a low mass collapsed object could  form. 
One possibility is such near-solar mass objects are primordial black holes (PBHs; \cite{carr}), although there are observational constraints on PBHs \cite{ck}.
An alternative is the capture of a PBH  
by a stellar host could transmute the host into a low mass black hole \cite{geno}.
As the probability of such a capture is small \cite{abra}, one needs to look for further alternatives.
According to one such scenario, dark matter (DM) particles could accumulate around the center of a neutron star and form a mini BH by collapse, which consumes the whole star by accreting matter from it \cite{gold}.
Thus, the whole neutron star is transmuted into a low mass BH. 

In general, a collapsed object is considered to be a BH due to the cosmic censorship conjecture (CCC) proposed by Penrose \cite{penrose}. According to the CCC, the nature abhors the singularity. Otherwise, one can see the singularity from infinity, in principle, and such a collapsed object is called as a naked singularity. It was shown \cite{jmn11} that the initial condition of gravitational collapse determines the nature of the end product (BH or naked singularity) after completion of the gravitational collapse. For example, if the collapse is initiated by a completely homogeneous pressureless dust, with the density being constant from the center to the boundary of the collapsing cloud, the singularity will be hidden inside the event horizon, and the end product will be a BH. On the other hand, if the gravitational collapse is induced by an inhomogeneous dust, the end product can be visible as a naked singularity. 
Moreover, for an inhomogeneous collapsing cloud with a non-zero pressure, a naked singularity can arise \cite{jmn14}. Note that, as a perfectly homogeneous collapse is an idealistic scenario (e.g., stellar density changes in the radial direction), 
a gravitational collapse of a star may give rise to a naked singularity as the end product. 
In fact, Sgr A* is recently proposed as a 1st type of
Joshi-Malafarina-Narayan (JMN-1 \cite{jmn11}) naked singularity \cite{eht2022}. Earlier, it was shown that M87* could be a Kerr-Taub-NUT naked singularity \cite{gcyl} or a Kerr naked singularity with the value of dimensionless Kerr parameter within $4.5$ to $6.5$ \cite{gcyl}. Similarly, GRO J1655-40 could be a Kerr-Taub-NUT naked singularity \cite{cbgm, cbgm2}. In fact, it is theoretically shown in several papers \cite{psj, jmn14, ss, jm, gss, jsy, gh} that the naked singularities are viable end products of collapse. 
In a recent paper \cite{pns}, it is also indicated that many of the proposed primordial BHs could actually be primordial naked singularities. 
There are some effects already predicted \cite{ckp, ckj, jmn14} to distinguish a naked singularity from a BH, by which it may be possible to identify near- and sub-solar mass naked singularities in future. As of now, the existence of naked singularity has been hinted in a few cases as mentioned above, but there is no confirmation yet from the astrophysical observation.

The nature of DM is not yet known. The interaction between the DM particles and the regular baryonic matter is via gravity, although there could be an extremely small additional interaction (with a very small cross section).
There are several proposed DM candidates and some of them could be asymmetric or non-self-annihilating (see figure 1 of \cite{gel} for details). Let us consider that 
the non-self-annihilating DM particles enter a cosmic object/host (e.g., neutron star, white dwarf, etc.). They lose energy due to the scattering with the baryonic matter of the host. For example, in case of the neutron star, the DM particles lose energy due to the scattering with the neutron. Now, the captured DM particles are thermally distributed in a very small region of the core of the host due to the large baryon density \cite{mcd}. Joining of the DM particles in an isothermal sphere around the core is known as thermalization \cite{dgr}. 
For neutron stars, the DM reaches thermal equilibrium with the neutrons, and the captured DM particles quickly occupy a very small region near the core of the host\footnote{We refer to sections III and IV of \cite{mcd} for more details.}.
If a sufficiently large number of DM particles accumulate at the core of the host, a tiny collapsed object may form by the  gravitational collapse \cite{dasg}. 
In all the papers (e.g., \cite{gold, dasg, dgr, mcd}, and references therein)  related to the dark core collapse, to the best of our knowledge, it has been presumed that the tiny collapsed object should be a BH, which may not be true. 
In this paper, we propose that depending on the initial condition (or, the nature) of the DM particles cloud, the tiny collapsed object could be either a BH or a naked singularity.
For example, if the accumulated DM cloud is inhomogeneous in nature with a non-zero, tangential pressure, an object-like endoparasitic naked singularity\footnote{Although this naked singularity is primarily hidden inside a stellar host, we call it a naked singularity in this sense that the singularity is not hidden inside the event horizon. For an object-like naked singularity, the event horizon does not form at all.} could form at the core of the stellar host. 
The endoparasitic naked singularity should start accreting matter from its host immediately after its formation, and transmute the entire host into a naked singularity. 
Note that it is recently proposed that the gravitational collapse also gives rise to the primordial naked singularities \cite{pns} instead of primordial BHs depending on the nature of the collapsing matter. 
If such a primordial naked singularity is captured by a stellar object, it may transmute into a collapsed object.
However, here we discuss only the transmutation led by an endoparasitic naked singularity formed by the accumulation of DM particles.

This paper is organized as follows. We discuss the Chandrasekhar limit in section \ref{s2}. section \ref{s3} describes the formation of an endoparasitic collapsed object inside a neutron star. 
We show in section \ref{s4} that the endoparasitic collapsed object could be a BH or a naked singularity depending on the initial conditions of the DM cloud. 
In section \ref{s5} we discuss the transmutation timescale, and we argue on the nature (BH or naked singularity) of the transmuted collapsed object in section \ref{s5.1}. 
We constrain a parameter space using the `missing pulsar' problem in section {\ref{mp}}, and conclude in section {\ref{s6}}.

\section{\label{s2}Chandrasekhar limit}
Here we review the derivation of Chandrasekhar limits for a system of fermions and bosons. Let us consider that $N$ number of fermions of mass $m$ are distributed over a sphere of radius $R$. Thus, the number density ($n$) of fermion is $n \sim N/R^3$, and the volume per fermion is $1/n$ due to the Pauli exclusion principle. The momentum of a fermion is obtained as $\sim \hbar n^{1/3}$ by using the Heisenberg uncertainty principle. The Fermi energy of a fermion in the relativistic regime is \cite{st}
\begin{eqnarray}
 E_{\rm F} \sim \hbar cn^{1/3} \sim \hbar c\f{N^{1/3}}{R}
\end{eqnarray}
whereas, the gravitational energy per fermion is
\begin{eqnarray}
 E_{\rm G} \sim -\f{GNm^2}{R}.
\end{eqnarray}
Here, $G$ is the Newton's constant, $\hbar$ is the Planck constant and $c$ is the speed of light in vacuum.
The total energy is 
\begin{eqnarray}
 E=E_{\rm F}+E_{\rm G} = \hbar c\f{N^{1/3}}{R} -\f{GNm^2}{R}.
\end{eqnarray}
If the total number of the fermions increases beyond \cite{ruf}
\begin{eqnarray}
 N_{\rm Ch}^{\rm f} \sim \left(\f{\hbar c}{Gm^2}\right)^{3/2}=\left(\f{M_{\rm P}}{m}\right)^{3}=1.8 \times 10^{51} \left(\f{100~{\rm GeV}}{m}\right)^3
  \label{f},
\end{eqnarray}
(where $M_{\rm P}=(\hbar c/G)^{1/2}=1.22 \times 10^{19}$ GeV is the Planck mass) the sign of $E$ becomes negative. This indicates that no equilibrium exists and gravitational collapse sets in.

The bosonic system is sigficantly different from the fermionic one as it has no Fermi pressure to prevent gravity. Since the bosons are confined inside a sphere with radius $R$, they have zero point energy $\hbar c/R$ for the Heisenberg uncetainty principle in the relativistic limit \cite{gle}. Thus, the typical energy for a boson in a sphere of radius $R$ is
\begin{eqnarray}
 E= \hbar c\f{1}{R} -\f{GNm^2}{R}.
\end{eqnarray}
In this case, the Chandrasekhar limit is obtained as \cite{ruf}
\begin{eqnarray}
  N_{\rm Ch}^{\rm b} \sim \left(\f{\hbar c}{Gm^2}\right)=\left(\f{M_{\rm P}}{m}\right)^{2} =1.5 \times 10^{34} \left(\f{100~{\rm GeV}}{m}\right)^2
\label{b}.
\end{eqnarray}
For both fermions and bosons, the gravitational collapse eventually leads to form a collapsed object.
Comparing eqs. (\ref{f}) and (\ref{b}), one can see that bosons usually collapse more easily than fermions.

\section{\label{s3}Formation of Endoparasitic Collapsed Objects inside Neutron stars}
Let us now consider that a neutron star of mass $M_{\rm h}$ and radius $R_{\rm h}$ begins to capture the asymmetric or non-self-annihilating dark matter (DM) particles at time $t=0$ with the capture rate ($C$) \cite{bl, mcd}
\begin{eqnarray}
 C = \sqrt{\f{6}{\pi}}\f{\rho_{\rm DM}}{0.45 m_{\c}} {2GM_{\rm h}R_{\rm h}\over \bar v_{\c}} . 
\left[1-\exp{\left(-\f{\s}{\s_{\rm cr}}\right)}\right] \left[1-\f{1-\exp{(-\xi)}}{\xi}\right]
\label{dmc}
\end{eqnarray}
with
\begin{eqnarray}\nonumber
 \xi=\f{6v_{\rm esc}^2}{{\bar v_{\c}}^2}\f{m_{\c}m_{\rm n}}{(m_{\c}-m_{\rm n})^2}
\end{eqnarray}
(see \cite{ps, mcd} for derivation), where $\rho_{\rm DM}$ is the density of the DM, $m_{\c}$ is the mass of a DM particle, $m_{\rm n}$ is the mass of one baryon of the host, $\bar v_{\c}=220$ km/s ($\equiv 10^{-3}c$) is the DM velocity dispersion, and $v_{\rm esc}=\sqrt{2GM_{\rm h}/R_{\rm h}}$ is the escape velocity from the neutron star. 
$\s/\s_{\rm cr} \leq 1$ indicates the efficiency factor for a given DM interaction cross section $\s$ \cite{kou, bl}. $\s_{\rm cr}= m_{\rm n} R_{\rm h}^2/(0.45 M_{\rm h})=10^{-45} {\rm cm}^2$ (where $m_{\rm n}=0.939$ GeV is the mass of a neutron) is the critical cross section \cite{kou} above which 
the DM particles passing through the neutron star are captured \cite{tak}.

We assume that the neutron star continuously captures the DM particles (fermions or bosons) from its surrounding  immediately after its formation. 
A tiny collapsed object could form when the DM particles satisfy the collapse criterion \cite{mcd, dasg}: 
\begin{eqnarray}
 N \geq {\rm Max}[N_{\rm Ch}, N_{\rm self}],
 \label{nc}
\end{eqnarray}
where $N$ is the total number of DM particles accumulated inside the neutron star. $N_{\rm Ch}$ (see eqs. \ref{f} and \ref{b}) and $N_{\rm self}$ are the Chandrasekhar limit and the number of DM particles required for initiating the self-gravitating collapse, respectively \cite{mcd}. 
If the total number ($N$) of  fermions(f)/bosons(b) increases beyond $N^{\rm f}_{\rm Ch}$ or $ N^{\rm b}_{\rm Ch}$ \cite{mcd},
the gravitational energy dominates over the total particle energy, and, the gravitational collapse occurs to form a tiny collapsed object in the core of the neutron star, under the condition of feeble repulsive DM self-interaction strength \cite{dgr}.  
Considering the scenario related to the Chandrasekhar limit (eq. \ref{f} and eq. \ref{b}), the bosonic DM particles should typically achieve the collapse criterion more easily than the fermionic ones, as mentioned in section \ref{s2}. 
However, this is not true for all the cases. 
The reason is, we also need to take into account the effect of $N_{\rm self}$ (see eq. \ref{nc}) which is as follows. Let us consider, a total DM mass $M_{\rm X}=N_{\rm X} m_{\chi}$ (where $N_{\rm X}$ is the total number of accumulated DM particles) is accumulated within a thermal radius of \cite{mcd}
\begin{eqnarray}
 r_{\rm th}=\left(\f{9k\T}{4\pi G\rho_{\rm B} m_{\chi}}\right)^{1/2},
 \label{rth}
\end{eqnarray}
where $\T$ is the core temperature of neutron star and $\rho_{\rm B}$ is the
baryon density. If the DM density\footnote{Note that here, for the sake of simplicity, $N_{\rm self}$ is derived assuming that the DM density and baryon density are uniform (see eqs. \ref{rth}-\ref{self}). 
The error introduced due to this assumption for a tiny region at the neutron star center is expected to be small.} 
is larger than the baryon density ($\rho_{\rm B}$) within the thermal radius ($r_{\rm th}$) \cite{mcd} of a host, i.e., 
\begin{eqnarray}
 \f{3M_{\rm X}}{4\pi r_{\rm th}^3} \gtrsim \rho_{\rm B} ,
 \label{rho}
\end{eqnarray}
the DM particles become self-gravitating.
Substituting eq. (\ref{rth}) in eq. (\ref{rho}) we obtain
\begin{eqnarray}
N_{\rm X} \equiv  N_{\rm self} \gtrsim \f{9}{2} \left[\f{1}{\pi \r_B m_{\chi}^5}\left(\f{k\T}{G}\right)^3 \right]^{1/2}
\label{self}
\end{eqnarray}
where, $k$ is the Boltzmann constant.
This means that the DM becomes self-gravitating once the total
number of accumulated DM particles is larger than the critical number $N_{\rm self}$. 
In this case, one has to consider eqs. (\ref{f}, \ref{b}, \ref{self}) carefully, and use eq. (\ref{nc}) to proceed further, because the `Max' function plays a crucial role here. However, for a typical neutron star, eq. (\ref{self}) reduces to 
\begin{eqnarray}
 N_{\rm self}=4.8 \times 10^{41} \left(\f{100~{\rm GeV}}{m_{\chi}}\right)^{5/2}\left(\f{\T}{10^5~{\rm K}}\right)^{3/2}
 \label{nself}
\end{eqnarray}
where the temperature of the core of the neutron star is taken in the order of $10^5$ K \cite{mcd}. We do not consider the possibility
of Bose-Einstein condensate (BEC) formation here, as the temperature for BEC formation is much less \cite{dasg} than the core temperature ($10^5$ K) of a neutron star.
Now, if the captured DM particles satisfy the above-mentioned collapse criterion (eq. \ref{nc}), a tiny endoparasitic collapsed object forms at the core of the neutron star. 
The mass of the newly formed collapsed object should be 
\begin{eqnarray}
 M_0=m_{\c}{\rm Max}[N_{\rm Ch}, N_{\rm self}] << M_{\rm h}
 \label{M0}
\end{eqnarray}
due to the accumulation of boson (or, fermion) DM particles. One can finally calculate the formation timescale ($t_0$) of the said collapsed object as 
\begin{eqnarray}
 t_0=t_{\rm th}+\f{N}{C}
 \label{t0}
\end{eqnarray}
where the thermalization timescale $(t_{\rm th})$ for a collapsed object inside a neutron star can be estimated by using the following expression \cite{mcd}:
\begin{eqnarray}
 t_{\rm th} \simeq 0.054~{\rm yr} \left(\f{m_{\chi}}{100~{\rm GeV}}\right)^2 \left(\f{2.1 \times 10^{-45}~{\rm cm}^{-2}}{\s}\right) \left(\f{10^5~{\rm K}}{\T}\right)
\end{eqnarray}
for $m_{\chi} \gtrsim 1$GeV.

\section{\label{s4}Nature of the Endoparasitic Collapsed Object}
In this section, we argue that the tiny endoparasitic collapsed object is not necessarily a BH, as assumed in almost all papers (e.g., \cite{gold, mcd, dasg, dgr}). 
The collapsed object could be a BH or a naked singularity depending on the density and pressure of the accumulated DM particles cloud, i.e., the initial condition of the collapsing material.
As mentioned in section~\ref{s1}, a naked singularity may arise due to an inhomogeneous collapse and non-zero pressure. 
It was also noted that a homogeneous collapse is an idealistic scenario. 
The actual process of collapse depends on many factors, such as how the captured DM particles accumulate at the stellar center after attaining the thermal equilibrium.

Let us now discuss if the DM can have pressure. 
It is not  clear what constitute the DM in galaxy halos. There are many possibilities as shown in figure 1 of \cite{gel}. 
One belief is that the DM is pressureless (cold dark matter) \cite{bk}. 
However, there is no direct detection of the cold dark matter particles yet. 
On the other hand, the flat rotation curves observed in the outer parts of the spiral galaxies can be explained well by a halo of DM with anisotropic pressures \cite{bk}. 
In fact, the bending of light rays passing through the DM halo is shown to be highly sensitive to the DM pressure (see \cite{bk} and references therein). 
DM pressure may play a significant role in some other cases as well (see \cite{fab, luo, bos21, rez, kur}). 
Intriguingly, it is also argued that dark energy may emerge from DM pressure \cite{cap, bos19, dag} at a cosmological level. 
These motivate us to consider a general scenario in this paper, i.e., the pressure of the DM particles cloud can be non-zero. 
Furthermore, since the DM pressure could also be anisotropic \cite{bos21, kur, bos22, bk, rez}, the 
collapse of the DM cloud can be anisotropic. 
In this context, we specifically refer to \cite{ruf} (see also \cite{gle, jet}) for an interesting and detailed discussion on the pressure of bosonic matter (and the fermionic matter) and its anisotropy.
Thus, the end product of the collapse of DM cloud at the center of a star can be a naked singularity.

\subsection{\label{s4a}DM particles cloud modeled as a perfect fluid or inhomogeneous dust}
Considering the above discussions, let us consider a general situation following \cite{pns} with the geometrized unit $(G=c=1)$. 
The components of the stress-energy tensor ($T^{\a}_{\b}$) of the DM cloud (fermions or bosons) in the comoving coordinates $(t,r,\theta,\phi)$ are given by
\begin{equation}
    T^{\a}_{\b}={\rm diag}\left(-\rho,p,p,p\right),
\end{equation}
where $\rho=\rho(t,r)$ and $p=p(t,r)$ are the density and the isotropic pressure of the collapsing DM cloud, respectively. The corresponding spacetime, i.e., the Lemaitre-Tolman-Bondi (LTB) metric is given by,
\begin{equation}
ds^2 = -dt^2 + R^{\prime 2}dr^2 + R^2 d\Omega ^2, \label{m1}
\end{equation}
where $d\Omega^2$ is the line element of the two-sphere at $R=R(t,r)$. Thus, one obtains \cite{jmn14}
\begin{equation}
\rho = \frac{F^{'}}{ R^{'} R^2}\label{denspres}
\end{equation}
and
\begin{equation}\label{efep}
     p= - \frac{\dot{F}}{\dot{R} R^2},
\end{equation}
where $F(r,t) = \dot{R}^2 R$ is the Misner-Sharp mass function 
of the collapsing DM cloud giving total mass within a coordinate radius $r$. Note that the dot denotes the partial derivative with respect to time coordinate $t$ and the prime
denotes the partial derivative with respect to the radial coordinate $r$. Actually, the collapsing spherical DM cloud consists of the concentric spherical shells, each of which is identified by the radial coordinate $r$.

For simplicity, if the collapsing scenario of the DM dust $(p=0$ and $F=F(r)$ \cite{dwi}) with a non-uniform density is considered, one obtains 
\begin{equation}
R(t,r)  = \left( r^\frac{3}{2} - \frac{3}{2} \sqrt{F(r)}t\right)^\frac{2}{3}, \label{physicalr}
\end{equation}
which can be rewritten (by defining a scaling function $s(t,r)=R(t,r)/r$) as 
\begin{equation}\label{timecurve}
    t(r,s)=\frac{2r^{3/2}}{3\sqrt{F(r)}}\left(1-s^{3/2}\right)
\end{equation}
to obtain the time curve. As the initial data is evolved, the physical radius of a spherical shell of fixed radial coordinate decreases and becomes zero at the singularity. The corresponding comoving time $t_v(r)$ is obtained as
\begin{equation}
t_{v}(r) =  \frac{2r^{\frac{3}{2}}}{3\sqrt{F(r)}}
\end{equation}
 by substituting $R(t_v,r)=0$ in eq. (\ref{physicalr}) or $s=0$ in eq. (\ref{timecurve}).
This, in fact, gives the singularity curve which depicts the spacetime singularity formation as the collapse endstate.
Note that the collapsing DM fluid spacetime (eq. \ref{m1}) is matched smoothly with the exterior Schwarzschild spacetime.
However, the singularity formed in the DM dust collapse scenario is event-like singularity.
This means that the singularity initially becomes visible in principle, as the event horizon does not form. In a later stage, the singularity gets covered within the event horizon, provided that the energy conditions ensuring the positivity of mass and energy density are satisfied. 
Thus, the conclusion is, the formation of a tiny endoparasitic BH could not be avoided eventually for an inhomogeneous dust collapse. 
The similar situation could arise, even if the DM cloud is modeled as a perfect fluid, and, hence, the conclusion remains same.

\subsection{\label{s4b}DM particles cloud modeled as an anisotropic fluid}
The following spherically symmetric metric \cite{jmn11}
\begin{equation}\label{m2}
    ds^2=-e^{2\nu}dt^2+\frac{R'^2}{H}dr^2+R^2d\Omega^2 \; 
\end{equation}
describes a dynamical gravitational collapse,
where $\nu$, $R$ and $H$ are functions of the comoving time $t$ and
the comoving Lagrangian radial coordinate $r$. 
Let us now consider that the DM particles cloud (fermions or bosons)
having only the tangential pressures ($p_\theta$), and is
described by the energy-momentum tensor as \cite{jmn11}
\begin{eqnarray}
 T^0_0=\rho, \; T_1^1=0, \; T_2^2=T_3^3=p_\theta. 
 \label{tmn}
\end{eqnarray}
The relation between the metric functions with the density and pressure 
is obtained from the Einstein equation.
In this model, we do not consider the
radial pressure contributions in the sense that the DM
pressure could be small enough to be that of dust, and the presence of tangential
pressures $(p_\theta)$ mimics the angular motion of the DM envelope while we are still considering a static model \cite{bos22}. Einstein introduced the non-vanishing tangential pressures to
describe a spherical and stationary configuration of collisionless
particles spinning under their mutual gravitational attraction \cite{ein}. 
Einstein's original assumption of all particles moving along circular paths around the center of symmetry of the cluster implies a condition on the geodesics of particles. When it is extended to a continuous distribution of particles, it determines the metric functions of the line element \cite{bos22}. In this sense, the metric mimics a cluster of spinning particles. All these above-mentioned properties (including the non-vanishing tangential pressure) of the DM particles should remain intact, even if they are accumulated in the core of the neutron star. 

Eqs. (\ref{m2}) and (\ref{tmn}) reveal that one obtains six unknowns ($\rho, p_{\th}, \nu, H, R$ and the mass function $F$ for the DM cloud) that appear due to the integration of the Einstein equations. As there are four Einstein equations, we are free to choose two functions. 
The free functions are chosen to be the mass function $F(r)$, which is the initial mass that is conserved for the DM cloud, and the tangential pressure of it. 
Although $F(r)$ is shown to be independent of time \cite{jmn11}, the tangential pressure depends on $r$ and $t$ via \cite{jmn11}
$v(r,t)$ as $p_\theta \equiv p_\theta(r,v)$. 
A global choice for the pressure
at all times fully fixes the evolution of the system.

Let us choose $F=M_{\rm e}r^3$ following \cite{jmn11}, that corresponds to a DM cloud initially homogeneous with a constant density. A scaling degree of freedom available in the definition of the physical radius $R$, using which one can write 
a scaling function $v(r,t)$ defined by $$R(r,t)=r\,v(r,t), \qquad v(r,t_i)=1.$$ The latter condition states that $R=r$ at the initial time $t=t_i$ from which the collapse develops. In the final equilibrium, i.e., in the static limit where the collapse stabilizes, one requires both the velocity and the acceleration of the in-falling shells to go to a vanishing value as the collapse progresses in future. This gives the limiting conditions \cite{jmn11}: $$\dot{v}=\ddot{v}=0.$$ 
We now have the following choices \cite{jmn11} (as mentioned above)
\begin{equation}
F(r)=M_{\rm e}r^3, \quad v_{\rm e}(r)=r^\alpha, \quad F(R)=M_{\rm e}R^\frac{3}{\alpha+1},
\label{toy}
\end{equation}
which imply an equilibrium state for the evolving collapse configuration, where the collapse of the DM cloud from a regular initial state settles to a final static solution which is an object-like naked singularity. Solving Einstein equations with the above initial conditions and the equilibrium metric in the interior
$R<R_{\rm b}$ with a central naked singularity is given by the JMN-1 spacetime \cite{jmn11}
\begin{equation}\label{m3}
ds^2_e =  -(1-M_{\rm e})\left(\frac{R}{R_{\rm b}}\right)^{\frac{M_{\rm e}}{1-M_{\rm e}}}dt^2+
\frac{dR^2}{1-M_{\rm e}}+R^2d\Omega^2 \; ,
\end{equation}
where the positive constant $M_{\rm e}$ should always be less than $1$ and 
$R_{\rm b}$ is the radius of the boundary of the JMN-1 spacetime where it is smoothly matched with the external Schwarzschild spacetime. We thus have a one-parameter family of static equilibrium solutions
parametrized by $M_{\rm e}$. Each member of this family of solutions has a naked singularity at the center \cite{jmn11}. 

To fix the behavior of $p_\theta$ at equilibrium, one needs to give the explicit form of the functions $H$ and $H_{,v}$ at equilibrium, and
both the equilibrium conditions $\dot{v}=0$
and $\ddot{v}=0$, which are necessary and sufficient for
this purpose \cite{jmn11}.  We have now freedom to choose the tangential
pressure in the evolving collapse phase which allows for the
different effective potentials to have different equilibrium
configurations. As the equilibrium configuration has a singularity at
the center, achieved as the limit of gravitational collapse,
one need to check under what conditions the singularity
will not be covered by the event horizon. From the
boundary condition at equilibrium, the condition is obtained as
$F/R<1$ \cite{jmn11}. In the final equilibrium state, one can write
$H$ in terms of energy density and pressure \cite{jmn11}, i.e.,
\begin{equation}
    \frac{F}{R_{\rm e}}=1-H_{\rm e}=\frac{4p_{\theta \rm e}}{\rho_{\rm e}+4p_{\theta \rm e}} \; .
\end{equation}
This is clearly smaller than unity for positive energy density ($\rho_{\rm e}$) and positive $p_{\theta \rm e}$. Finally, we obtain the interesting result that any central singularity that forms in the final equilibrium
configuration via the collapse dynamics described here is
always a naked singularity, i.e., the singularity is not covered by the event horizon.

\section{\label{s5}Transmutation timescale of the host neutron star}
In section \ref{s4}, we show that the endoparasitic collapsed object could be a BH (see section \ref{s4a}) or a naked singularity (see section \ref{s4b}) depending on the initial conditions. However, after its formation it will start to accrete matter immediately from its host neutron star with the spherical Bondi accretion rate: 
\begin{eqnarray}
 \f{dm}{dt}\Bigg|_{\rm acc}=k_1 m^2 \, ,
 \label{acc}
\end{eqnarray}
where $k_1=3 G^2 M_{\rm h}/(c_{\rm s}^3 R_{\rm h}^3)$ \cite{geno, bs} and $c_{\rm s}$ is the speed of sound ($c_{\rm s} \sim 0.9c$) inside the neutron star. 
\begin{figure}
 \begin{center}
{\includegraphics[width=3in,angle=0]{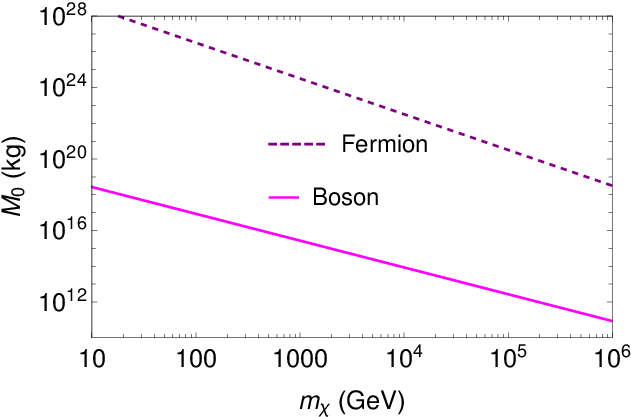}}
\caption{\label{fm0}Mass ($M_0$) of the endoparasitic collapsed object (BH or naked singularity) formed inside a neutron star of $M_{\rm h}=1.4M_{\odot}$ at $t=t_0$ versus the mass ($m_{\chi}$) of DM particles. This figure explains that $M_{\rm h} >> M_0$. See section \ref{s5} for details.}
\end{center}
\end{figure}
Now, by integrating eq. (\ref{acc}):
\begin{eqnarray}\nonumber
 \int^{t_{\rm d}}_{t_0} dt & = & \int_{M_0}^{M_{\rm h}} \f{dm}{k_1 m^2}
 \\
 \Rightarrow t_{\rm d} - t_0 &=& \f{1}{k_1 M_0}\left[1-\f{M_0}{M_{\rm h}} \right],
 \label{tdp}
 \end{eqnarray}
the transmutation timescale $(t_{\rm d})$ of the neutron star is obtained as
 \begin{eqnarray}
\Rightarrow t_{\rm d} & \approx & t_0 + \f{c_{\rm s}^3 R_{\rm h}^3}{3 G^2 M_{\rm h} M_0}
\equiv  t_0+t_{\rm acc}
 \label{td}
\end{eqnarray}
where, $t_{\rm acc}$ is the accretion timescale after the formation of endoparasitic collapsed object. To obtain eq. (\ref{td}) from eq. (\ref{tdp}), we neglect the term $M_0/M_{\rm h}$ as $M_{\rm h} >> M_0$ which is clear from figure \ref{fm0}. The solid magenta and dashed purple curves of figure \ref{fm0} indicate the mass ($M_0$) of the newly-formed endoparasitic collapsed object inside a typical neutron star of $M_{\rm h}=1.4M_{\odot}$ by collapsing of particles, respectively.
Note that the Bondi (spherical) accretion rate (eq. \ref{acc}) should be same \cite{chak} for a BH and a naked singularity. Since this should also be true for eq. (\ref{td}), we only need to know $M_{\rm h}$ and the mass of the endoparasitic collapsed object $M_0$ (see eq. \ref{M0}) to calculate $t_{\rm d}$, regardless of its nature.

\begin{figure}
 \begin{center}
\subfigure[$~ \rho_{\rm DM} \sim 0.4$ GeV.cm$^{-3}$ ($D \sim 10$ kpc)]{\includegraphics[width=2.6in,angle=0]
{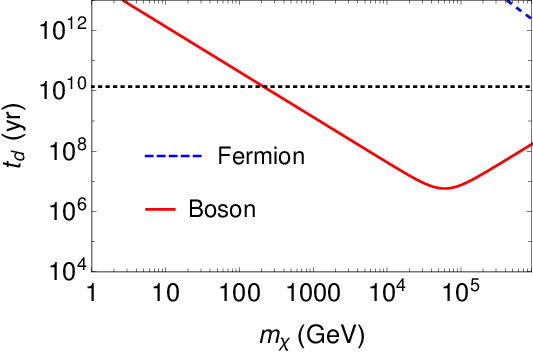}}
\hspace{0.05\textwidth}
\subfigure[$~ \rho_{\rm DM} \sim 7 \times 10^4$ GeV.cm$^{-3}$ ($D \sim 0.1$ pc)]
{\includegraphics[width=2.6in,angle=0]{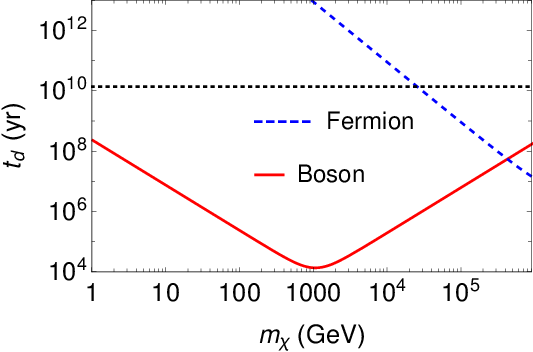}}
\hspace{0.05\textwidth}
\caption{\label{ftd}The variation of transmutation timescale ($t_{\rm d}$) of a neutron star of $M_{\rm h}=1.4M_{\odot}$ and $R_{\rm h}=10$ km as a function of the mass ($m_{\chi}$) of non-self-annihilating DM particles. Panels (a)-(b) correspond to two different DM density profiles with a fixed DM particle-nucleon scattering cross section, i.e., $\s > 10^{-44}$ cm$^2$. The black dotted horizontal line represents the Galactic age ($t_{\rm Ga}$). This figure shows that there is a non-zero possibility of transmutation of the neutron star into a collapsed object (BH or naked singularity depending on the initial conditions) in a reasonable time for a wide range of $m_{\chi}$ values, depending on the value of DM density profile ($\rho_{\rm DM}$). See section \ref{s5} for details.}
\end{center}
\end{figure}

\begin{figure}
 \begin{center}
 \subfigure[$~\rho_{\rm DM} \sim 0.4$ GeV.cm$^{-3}$ ($D \sim 10$ kpc)]{\includegraphics[width=2.6in,angle=0]
{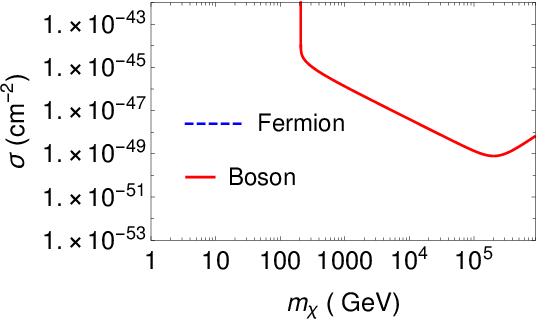}}
\subfigure[$~\rho_{\rm DM} \sim 7 \times 10^4$ GeV.cm$^{-3}$ ($D \sim 0.1$ pc)]{\includegraphics[width=2.6in,angle=0]
{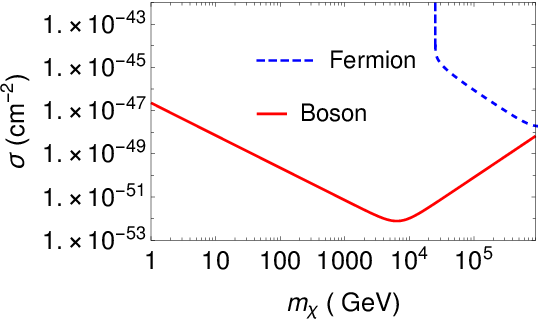}}
\hspace{0.05\textwidth}
\caption{\label{fps}Constraints on the DM particle-neutron scattering cross section $\s$, as a function of mass ($m_{\chi}$) of the DM particles for the possible transmutation of a typical neutron star, if it is located at a distance $D$ with the specified $\rho_{\rm DM}$ value mentioned in the caption of each panel. The parameter space located on the right side of the red/blue-dashed curve indicates a transmutation for $t_{\rm d} < t_{\rm Ga}$. On the other hand, the left side of the red/blue-dashed curve suggests that the transmutation may occur in the future.  See section \ref{s5} for details.
}
\end{center}
\end{figure}

For the purpose of demonstration to calculate $t_{\rm d}$,
let us assume that a DM particle is captured after a single scattering, and, hence, we consider here the non-self-annihilating DM particles with a mass range: $1$ GeV $< m_{\chi} < 10^6$ GeV \cite{dgr}. 
If a typical neutron star of $M_{\rm h} \sim 1.4M_{\odot}$ and $R_{\rm h} \sim 10$ km captures the DM particles in the above mass range and forms an endoparasitic BH/naked singularity in its core, the entire neutron star could transmute into a BH/naked singularity depending  on the initial condition of collapse as discussed in section \ref{s4}. 
Now, if the transmutation time ($t_{\rm d}$) of a neutron star satisfies the following condition: $t_{\rm d} < t_{\rm Ga}$ (where $t_{\rm Ga} \sim 1.36 \times 10^{10}$ yr is the Galactic age), the neutron star could have already transmuted. Note that $t_{\rm d}$ depends on the DM density ($\rho_{\rm DM}$) profile (see eq. \ref{dmc}, eq. \ref{t0} and eq. \ref{td}) as well. Thus, for the purpose of demonstration, here we consider the most commonly used Navarro-Frenk-White (NFW) DM density profile \cite{nfw},
\begin{eqnarray}
 \rho_{\rm DM} (D) &=& \rho_{0\rm DM} ~(D/D_s)^{-\a}(1+D/D_s)^{-3+\a} 
 \label{nfw}
\end{eqnarray}
where, $D$ is the distance from the Galactic center. We consider the local DM density $\rho_{0\rm DM}=0.4$ GeV.cm$^{-3}$ following \cite{nfw, jet}, the scale radius $D_s=20$ kpc following \cite{bl} with $\a=1$ \cite{bl}. Thus, $\r_{\rm DM}\sim 7 \times 10^4$ GeV.cm$^{-3}$ \cite{bl} is obtained at a distance $D \sim 0.1$ pc inside the Galactic center,  and $\r_{\rm DM}\sim 0.4$ GeV.cm$^{-3}$ is for the Galactic disk ($D \sim 10$ kpc) \cite{hoo}.
Substituting all the values in eq. (\ref{td}), one can plot $t_{\rm d}$ versus $m_{\chi}$ for the DM particle-neutron cross section $\s > 10^{-44}$ cm$^{-2}$ for the two different values of DM density. 

Panel (a) of figure \ref{ftd} shows that the transmutation of a neutron star could not be possible in the Galactic disk for the fermionic DM particles, whereas it could be possible for the bosonic DM particles. Panel (b) corresponds to the Galactic bulge, and it shows that the transmutation could be possible for both the fermionic and bosonic DM particles. However, comparing both the panels we can see that the possibility of transmutation within $t_{\rm d} \leq t_{\rm Ga}$ decreases with decrement of $\rho_{\rm DM}$. Note that figure \ref{ftd} stands for $\s > 10^{-44}$ cm$^{-2}$ only, and, thus it can be considered just as an example of how $t_{\rm d}$ varies with $m_{\chi}$ for three different $\rho_{\rm DM}$ values with a fixed value of $\s$. One cannot constrain the $\s - m_{\chi}$ parameter space from figure \ref{ftd}. Panels (a)-(b) of figure \ref{fps}, in fact, help to constrain the non-self-annihilating DM-neutron scattering cross section $\sigma$, as a function of $m_{\chi}$ for the possible transmutation within $t_{\rm d} \leq t_{\rm Ga}$, if the neutron star is located in the Galactic disk and Galactic bulge, respectively. The parameter space located on the right side of the red (blue-dashed) curve indicates a transmutation for $t_{\rm d} \leq t_{\rm Ga}$. On the other hand, the left side of the red (blue-dashed) curve suggests that the transmutation may occur in the future. 

\section{End state of the transmuted object\label{s5.1}}
Now, the question is, what would be the end state of the transmuted object. The tiny endoparasitic BH formed (as discussed in section \ref{s4a}) due to the collapse of a perfect DM fluid or dust should transmute the host into a BH. 
The transmutation process should be as discussed in various papers (e.g., \cite{mcd, dasg}). 
If the host neutron star has a non-zero spin (e.g., a pulsar), the transmutation process could be evolved as described in \cite{el}. However, the end state of the transmuted object would be a BH, and not a naked singularity as shown in \cite{cbns}.

When an endoparasitic naked singularity (as discussed in section \ref{s4b}) is formed due to an anisotropic DM fluid collapse, it starts to accrete matter from its host and eventually the entire host would transmute into a naked singularity.
As an exactly non-spinning object may not exist in the Universe \cite{jh}, the host neutron star should have a non-zero spin parameter $a_{\rm h}$ $\equiv J_{\rm h}/M_{\rm h}$, where $J_{\rm h}$ is the angular momentum of the host. Since the spin frequency ($\o$) of both the neutron star and the endoparasitic naked singularity during its formation (i.e., at ${t \ra t_0}$) should be the same (as $M_0 << M_{\rm h}$) \cite{cbns},
the endoparasitic naked singularity should also have a non-zero spin parameter. Let us denote that spin parameter of the endoparasitic naked singularity as $a_0$ at time $t \ra t_0$. However, if one calculates the numerical value of $a_0/M_0$, it comes as negligible small (as $M_0 << M_{\rm h}$), and, hence, one can consider it as $a_0 \ra 0$ for all the practical purposes.  During accretion from its host, the value of spin paramater starts to increase from $a_0$ and could eventually acquire $\sim a_{\rm h}$ \cite{cbns} at the end of the transmutation process (i.e., $t=t_{\rm d}$), if the dynamical ejecta is small \cite{el, cbns}. If the dynamical ejecta is large \cite{cbns} during accretion by the endoparasitic naked singularity, the final transmutation could give birth to a sub-solar-mass or near-solar mass slowly-spinning JMN-1 naked singularity. Computation of the evolution of such an endoparasitic naked singularity, which is a complex and entirely unexplored field, 
will be considered in future research works.

The slowly-spinning JMN-1 naked singularity spacetime is recently derived as \cite{bam}
\begin{eqnarray}\label{17}
ds^2=-(1-M_{\rm e})\left(\frac{R}{R_{\rm b}}\right)^\frac{M_{\rm e}}{(1- M_{\rm e})} dt^2+\frac{dR^2}{1-M_{\rm e}}+R^2\left(d\theta^2+\sin^2\theta d\phi^2\right) \nonumber \\
-2a\sin^2\theta\left[\left(\frac{R}{R_{\rm b}}\right)^{\frac{M_{\rm e}}{2(1- M_{\rm e})}}-(1-M_{\rm e})\left(\frac{R}{R_{\rm b}}\right)^{\frac{M_{\rm e}}{(1- M_{\rm e})}}\right]dt d\phi
\label{rJMN}
\end{eqnarray}
where, $a$ is the spin parameter of this spacetime. The above slowly-spinning JMN-1 spacetime (eq. \ref{rJMN}) can be matched with the exterior slowly-spinning Kerr spacetime at $R=R_{\rm b}$ \cite{bam}, 
\begin{eqnarray}
\label{eq:slowly_rotating_kerr}
ds^2 = -\left(1-\frac{M_{\rm e} R_{\rm b}}{R}\right) dt^2+\frac{dR^2}{\left(1-\frac{M_{\rm e} R_{\rm b}}{R}\right)} +R^2\left(d\theta^2+\sin^2\theta d\phi^2\right)
-\frac{2aM_{\rm e}R_{\rm b}}{R}\sin^2\theta ~dt d\phi\nonumber \\
\end{eqnarray}
where $2M=M_{\rm e} R_{\rm b}$ is the Schwarzschild mass of the slowly-spinning ($a/M << 1$) naked singularity. Eq. (\ref{rJMN}) reduces to eq. (\ref{m3}) for $a \ra 0$, which describes the non-spinning JMN-1 naked singularity spacetime.  

Our purpose here to introduce eq. (\ref{rJMN}) is that it could describe the spacetime of a slowly-spinning naked singularity, formed due to the transmutation of a spinning neutron star. For example, if the DM particles are accumulated inside a pulsar like the fastest known spinning pulsar PSR J1748-2446ad with $\o \sim 716$ Hz \cite{hes} (i.e., $ a_{\rm h}/M_{\rm h}\sim 0.09$), and it satisfies the collapse criterion, it could be transmuted into a slowly-spinning JMN-1 naked singularity of the same spin parameter (i.e., $ a/M \sim 0.09$), provided that the dynamical ejecta is negligible. Even for a typical rapidly-spinning neutron star, the maximum spin parameter range could be around $ a_{\rm h}/M_{\rm h} \sim 0.1$ (but see \cite{bha17}).
In such a scenario, one can study the evolution of the endoparasitic naked singularity and transmutation of a neutron star with the help of slowly-spinning JMN-1 naked singularity metric (eq. \ref{rJMN}) following \cite{el} instead of Kerr spacetime. However, this metric is not enough to study the evolution of an endoparasitic naked singularity formed inside a fast-spinning white dwarf \cite{cbns}, earth-like planet, sun-like star etc., as the spin parameter is very high for these objects \cite{cbns}.

\section{\label{mp}Constraining  $\s - m_{\chi}$ parameter space from the Missing pulsar problem}~

Here, we propose the following general technique to constrain the $\s - m_{\chi}$ parameter space from the missing pulsar problem.
Let us assume the following ideal situation.
\\
(1) Within a distance $\mathcal{D}$ around the Galactic center, all old pulsars (which are almost as old as the Galactic age $t_{\rm Ga}$) have been transmuted into BHs and/or naked singularities. 
Hence, no
such pulsar exists in this region.
\\
(2) Outside this region, no pulsar has been transmuted.

\noindent
Point (1) should give a lower limit of $\s$ 
and Point (2) should give an upper limit of $\s$, each as a function of $m_{\chi}, ~\mathcal{D}$, DM profile, particle type (boson or fermion), etc. 
Thus, the $t_{\rm d} = t_{\rm Ga}$ condition should ideally give a value of $\s$ as a function of $m_{\chi}, ~\mathcal{D}$, DM profile, particle type (boson or fermion), etc.
The transmutation timescale $t_{\rm d}$ can be calculated using eq. (\ref{td}) and eq. (\ref{t0}).

\begin{figure}
 \begin{center}
\subfigure[~NFW DM density profile (eq. \ref{nfw}): $\rho_{\rm DM}=800$ GeV.cm$^{-3}$]{\includegraphics[width=2.6in,angle=0]
{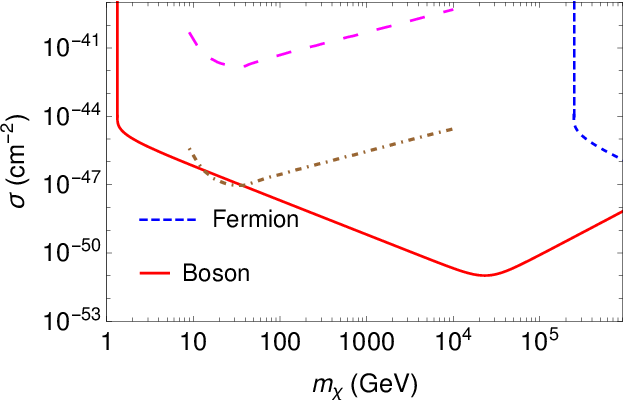}}

\hspace{0.05\textwidth}
\subfigure[~Cored DM profile (eq. \ref{core}): $\rho_{\rm DM} = 17$ GeV.cm$^{-3}$]
{\includegraphics[width=2.6in,angle=0]{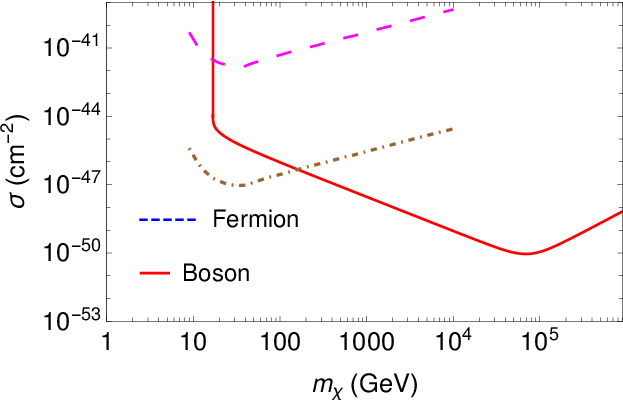}}
\hspace{0.05\textwidth}
\subfigure[~Cored DM profile (eq. \ref{core}): $\rho_{\rm DM} =410$ GeV.cm$^{-3}$]
{\includegraphics[width=2.6in,angle=0]{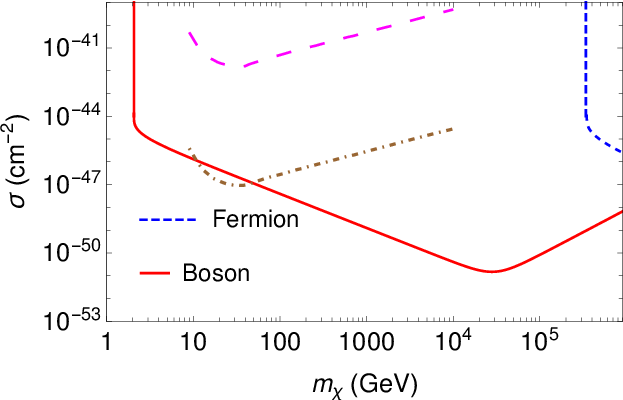}}
\caption{\label{fmpp}Constraints on the DM-nucleon scattering cross section $\s$, as a function of the PDM particle mass $m_{\chi}$ for the possible transmutation. The red-solid / blue-dashed curve corresponds to $t_{\rm d} = t_{\rm Ga}$ for boson/fermion, respectively, assuming all the old pulsars located within a distance $\mathcal{D} \sim 10$ pc around the Galactic center have been transmuted into BH or naked singularity. The parameter space located on the right side of the red-solid / blue-dashed curve suggests a transmutation for $t_{\rm d} \leq t_{\rm Ga}$, whereas the left side of the red-solid / blue-dashed curve suggests that the transmutation may occur in the future. The dot-dashed brown and dashed magenta curves (equivalent to solid black curves of figure 5 and figure 7 of \cite{aal}, respectively) are due to the constraints given by the LZ experiment for the nominal upper limit on the DM-nucleon spin-independent and DM-neutron spin-dependent cross section as a function of $m_{\chi}$, respectively. This figure shows the constrained $\s-m_{\chi}$ parameter spaces obtained due to the missing pulsar problem for the three different DM density profiles. See section \ref{mp} for details.}
\end{center}
\end{figure}

Now, let us consider a specific example with $\mathcal{D} \sim 10$ pc following \cite{ful, fer}.
We will consider the Point (1) mentioned above, but not the Point (2), to be conservative.
The results are shown in figure \ref{fmpp}.
The value of $\rho_{\rm DM}$ at a distance $\sim 10$ pc can be calculated to be 
$\sim 800$ GeV.cm$^{-3}$ assuming the NFW DM density profile (eq. \ref{nfw}).
The parameter space located on the right side of the red-solid (blue-dashed) curve of Panel (a) of figure \ref{fmpp} indicates a transmutation for $t_{\rm d} < t_{\rm Ga}$.
On the other hand, the left side of the red-solid (blue-dashed) curve of the same plot suggests that the transmutation may occur in the future. 
Thus, the Point (1) mentioned above and the DM density at $10$ pc give a lower limit of $\s$ as a function of $m_{\chi}$. 

The dot-dashed brown and dashed magenta curves in Panel (a) of figure \ref{fmpp} are due to the constraints given by the LUX-ZEPLIN (LZ 2022) experiment \cite{aal} for the nominal upper limit on the DM-nucleon spin-independent and DM-neutron spin-dependent cross section as a function of $m_{\chi}$, respectively. Note that the dot-dashed brown and dashed magenta curves in Panel (a) of figure \ref{fmpp} are equivalent to solid black curves of figure 5 and figure 7 of \cite{aal}, respectively. Since both the LZ \cite{aal} and XENONnT \cite{apr} DM search experiments have considered the DM particles mass range upto $m_{\chi} \sim 10^4$ GeV, we indicate the upper limit of $\s$ upto the same $m_{\chi}$ value.

Now, if the pulsars are missing from the region with distance $< 10$ pc \cite{ful, fer} in the Galactic center, the $\s - m_{\chi}$ parameter space bounded by the red-solid curve and the dot-dashed brown curve of Panel (a) of figure \ref{fmpp} could be considered as the suitable constrained parameter space for the bosonic DM particles obtained from the missing pulsar problem assuming the DM-nucleon spin-independent interaction for the NFW DM density profile.
Similarly, the $\s - m_{\chi}$ parameter space bounded by the red-solid curve and the dashed magenta curve of Panel (a) of figure \ref{fmpp} could be considered as the suitable constrained parameter space for bosonic DM particles obtained from the missing pulsar problem assuming the DM-neutron spin-dependent interaction.
However, since the blue-dashed curve is located far from the dot-dashed brown and dashed magenta curves, the upper limit could not be suggested from the LZ and XENONnT experiments for the fermionic DM particles.
Note that here we do not specifically consider a spin-independent or spin-dependent DM-nucleon interaction, but, by studying Panel (a), one can see that the possibility of the transmutation of neutron stars into BHs/naked singularities is more for the spin-dependent DM-neutron interaction.

Let us now consider a different type of DM density profile to check whether the transmutation in $t_{\rm d} \leq t_{\rm Ga}$ is possible at all, and, if yes, how the $\s-m_{\chi}$ parameter space depends on it. 
To check this, here we consider a cored DM density profile which is expressed in the Galactic plane as \cite{oll, cbns}:
\begin{eqnarray}
 \rho_{\rm DM} (D) &=& \rho_{0\rm c}/[1+(D/D_{\rm c})^2],
 \label{core}
\end{eqnarray}
where, $D_{\rm c}$ is the core radius and $\rho_{0\rm c}$ is the central density of the dark halo. 
We consider two different values from Table 2 of \cite{oll} following \cite{cbns} for the purpose of demonstration. The first one is $\rho_{0\rm c}=0.421 M_{\odot}$ pc$^{-3} \sim 17$ GeV.cm$^{-3}$ for $D_{\rm c}=1$ kpc \cite{oll}, and the second one is $\rho_{0\rm c} =9.884 M_{\odot}$ pc$^{-3} \sim 410$ GeV.cm$^{-3}$ for $D_{\rm c}=0.2$ kpc \cite{oll}. Panels (b) and (c) of figure \ref{fmpp} correspond to $\rho_{\rm DM} \sim 17$ and $\sim 410$ GeV.cm$^{-3}$, respectively. 
These panels are similar to Panel (a) of figure \ref{fmpp}. The blue-dashed curve is out of the range of Panel (b). 
Comparing Panels (a)-(c) of figure \ref{fmpp}, one can conclude that the area of constrained  $\s - m_{\chi}$ parameter space decreases with the decrease of $\rho_{\rm DM}$, as expected. 
Our result shows that there are ample possibilities of the transmutation of neutron stars into BHs/naked singularities, and the consideration of different density profiles does not change our conclusion.

Note that, while the above specific example is for $\mathcal{D} \sim 10$ pc, the technique may work for any region around the Galactic center without sufficiently old pulsars, if we assume that the lack of such pulsars are due to transmutation \cite{ful, bl, lav}.
Future observations (e.g., with Square Kilometre Array \cite{ska}) will have significant impact on this by constraining the pulsar population near the Galactic center.

\section{\label{s6}Discussion and Conclusion}

In this paper, we have shown that a dark core collapse could lead to not only a BH but also a naked singularity, depending on the initial condition of the DM cloud. 
A recent paper \cite{cbns} has shown that some white dwarfs can realistically transmute into Kerr BHs and Kerr naked singularities ($a/M > 1$) with a wide range of near- and sub-solar-mass values by capturing the asymmetric or non-self-annihilating DM particles.
But that paper \cite{cbns} has also shown that a neutron star, even if a rapidly spinning one, cannot transmute into a naked singularity with $a/M > 1$ by accretion.
On the other hand, the current paper, for the first time to the best of our knowledge, argues that depending on the initial conditions, an endoparasitic naked singularity
could form instead of an endoparasitic BH inside a neutron star, and the same neutron star could transmute into a naked singularity. 
Apart from different progenitors,
there is a conceptual difference between the two papers. 
In \cite{cbns}, the naked singularity is proposed to form due to the accretion by an endoparasitic BH. This is a Kerr superspinar. Thus, a neutron star cannot give rise to such a naked singularity (see Appendix A of \cite{cbns}), but a white dwarf can. 
Consequently, the mass of the naked singularity cannot be more than that of a white dwarf (implying mostly sub-solar or slightly super-solar mass). 
On the other hand, in the current paper, a naked singularity is formed directly by collapse and not from a BH. It is not a Kerr superspinar and hence a different kind of naked singularity from the former. 
This is the main conceptual difference. 
In this case, compared to the previous one, usually mass should be more but the spin should be less, when the collapse happens at the core of a neutron star.
Even if the end product is a BH, and not a naked singularity, the mass of the final BH for the former paper (i.e., \cite{cbns}) should typically be less than that for the current paper. These differences could also be helpful to determine whether the progenitor was a neutron star or a white dwarf. 
However, it is possible that 
a low-mass BH or a low-mass naked singularity with a low spin parameter value could form via the accretion by an endoparasitic collapsed object within either a white dwarf or a neutron star of similar masses and similarly low spin parameter values. 
In such a case, the progenitor cannot be easily identified either as a white dwarf or as a neutron star.
While a detailed study of the said accretion process might provide a way to identify the progenitor in this case, such a process is not yet well-studied, would involve detailed fluid dynamical computations, and is not within the scope of this paper.

We have shown that a neutron star could transmute into a JMN-1 naked singularity with $a/M < 1$. 
Intriguingly, JMN-1 naked singularity spacetime shows that a naked singularity could arise regardless its spin parameter value ($a$ could be zero or non-zero) unlike Kerr spacetime. 
Note that we have assumed the Bondi (spherical) accretion in this paper. Although the Bondi accretion is not exactly valid (as the matter of a host possesses a non-zero angular momentum), this assumption has no significant effect on the estimation of the accretion or destruction timescale \cite{kt, rbs} even for the rapidly spinning stars \cite{el, cbns}.

In this paper, we have considered the non-self-annihilating DM particles with a mass range: $1$ GeV $< m_{\chi} < 10^6$ GeV \cite{dgr, cbns}. This is sufficient for our purpose of studying the formation of low mass collapsed objects which could have been involved in GW190425 and GW190814. Our result is also commensurate with both the LZ \cite{aal} and XENONnT \cite{apr} DM search experiments, as seen from figure \ref{fmpp}. It is hardly possible to form such a low mass collapsed object by accumulating the DM particles of mass less than the MeV range within the time $t_{\rm Ga}$, as seen from figure \ref{ftd}. However, in a recent paper \cite{argu}, the supermassive black hole formation has been proposed from the gravitational collapse of fermionic dense DM cores of particles of keV mass range at the centre of DM haloes.

It was proposed \cite{ful, bl, lav} that the missing pulsars in the Galactic center region have already transmuted (within a period of time less than the age of the Galaxy) into BHs due to the dark core collapse initiated by the DM particles \cite{bl}. However, if the DM cloud is modeled as an anisotropic fluid, the missing pulsars in the Galactic center region could have transmuted into slowly-spinning JMN-1 naked singularities, and such singularities could be found abundantly near the Galactic center. 
Their existence in this region could also indirectly show that the DM is not pressureless.

\end{document}